\documentstyle{article}

\def\abstract#1{\vskip 7mm
        \begin{center}{\large Abstract}\par \smallskip
                \begin{minipage}[c]{12cm}
                        \small #1
                \end{minipage}
        \end{center}
}
\def\title#1{\begin{center}{\Large\bf #1}\end{center}}
\def\author#1{\vskip 5mm \begin{center}{#1}\end{center}}
\def\address#1{\begin{center}{\it #1}\end{center}}
\makeatletter
\@ifundefined{lesssim}{}{}
\@ifundefined{gtrsim}{}{}
\def\vereq#1#2{\lower3pt\vbox{\baselineskip1.5pt \lineskip1.5pt
\ialign{$\m@th#1\hfill##\hfil$\crcr#2\crcr\sim\crcr}}}
\makeatother

\begin{document}

\title{%
Doubly-gauge-invariant formalism of brane-world cosmological
perturbations 
}
\author{%
Shinji Mukohyama
}
\address{%
Department of Physics, Harvard University\\
Cambridge, MA 02138, USA
}
\abstract{
We review the doubly gauge invariant formalism of cosmological
perturbations in the Randall-Sundrum brane world. This formalism leads
to four independent equations describing the evolution of scalar 
perturbations. Three of these equations are differential equations
written in terms of gauge invariant variables on the brane only, and
the other is an integro-differential equation describing non-locality
due to bulk gravitational waves. At low energy the evolution of the
scalar-type cosmological perturbations in the brane-world cosmology
differs from that in the standard cosmology only by non-local effects
due to bulk gravitational waves. 
}


\section{Introduction}


The idea that our four-dimensional world may be a timelike surface, or
a world-volume of a $3$-brane, in a higher dimensional spacetime has
been attracting a great deal of physical interests. As shown by
Randall and Sundrum~\cite{RS2}, the $4$-dimensional Newton's law of
gravity can be reproduced on a $4$-dimensional timelike hypersurface
with positive tension in a $5$-dimensional AdS background despite the
existence of the infinite fifth dimension.


Moreover, cosmological solutions in the Randall-Sundrum brane world
scenario were found~\cite{BDEL,Mukohyama2000a,Kraus,Ida}. In these
solutions, the standard Friedmann equation is restored at low energy, if
a parameter in the solutions is small enough. If the parameter is not 
small enough, it affects cosmological evolution of our universe as
dark radiation~\cite{Mukohyama2000a}. Hence, the parameter should be
very small in order that the brane-world scenario should be consistent
with nucleosynthesis~\cite{BDEL}. On the other hand, in
ref.~\cite{MSM}, it was shown that $5$-dimensional geometry of all
these cosmological solutions is the Schwarzschild-AdS (S-AdS)
spacetime and that the parameter is equivalent to the mass of the black
hole. Therefore, the $5$-dimensional bulk geometry should be the S-AdS
spacetime with a small mass, which is close to the pure AdS
spacetime. Moreover, black holes with small mass will evaporate in a
short time scale~\cite{Hawking}. Thus, it seems a good approximation to
consider the pure AdS spacetime as a $5$-dimensional bulk geometry for
the brane-world cosmology.

For the AdS bulk spacetime, the brane world scenario can reproduce the
standard cosmology as evolution of a homogeneous isotropic universe at
low energy. Hence, this scenario may be considered as a realistic
cosmology and it seems effective to look for observable consequences of
this scenario. For this purpose, cosmic microwave background (CMB)
anisotropy is a powerful tool. Therefore, we would like to give
theoretical predictions of the brane-world scenario on the CMB 
anisotropy. There are actually many papers on this
subject~\cite{Mukohyama2000b,KIS,Mukohyama2000c,Maartens,Langlois,BDBL,Koyama-Soda,LMW,LMSW,Mukohyama2001}.
However, the calculation of the CMB spectrum is not an easy task. The
main difficult problems are the following two: (i) how to give the
initial condition; (ii) how to evolve perturbations. As for the first
problem, there is essentially the same issue even in the standard
cosmology. In this paper we shall concentrate on the problem (ii).


\section{Doubly gauge-invariant formalism}
	\label{sec:formalism}

In this section we review some important points of the gauge-invariant
formalism of gravitational perturbations in the bulk and the doubly
gauge-invariant formulation of the perturbed junction condition. For
details, see refs.~\cite{Mukohyama2000b,Mukohyama2000c}.

\subsection{Master equation in the bulk}

Now let us consider gravitational perturbations in $D$-dimensional
maximally-symmetric spacetimes since in the simple brane world scenario
the background bulk geometry is known to be an AdS spacetime, one of
three maximally symmetric spacetimes. Since a general motion of
homogeneous, isotropic $(D-2)$-brane breaks the symmetry of the
$D$-dimensional maximally symmetric spacetime to that of a
$(D-2)$-dimensional constant-curvature space, we consider the following
decomposition of the background spacetime. 
%
\begin{equation}
 g^{(0)}_{MN} = \gamma_{ab}dx^adx^b+r^2\Omega_{ij}dx^idx^j,
        \label{eqn:gamma-Omega}
\end{equation}
where $\Omega_{ij}$ is a metric of a ($D-2$)-dimensional
constant-curvature space, $\gamma_{ab}$ is a $2$-dimensional metric
depending only on the $2$-dimensional coordinates $\{x^a\}$, and $r$
also depends only on $\{x^a\}$. 

Now let us analyze metric perturbations around the maximally symmetric
background. First, let us expand metric perturbations by harmonics on
the ($D-2$)-dimensional constant-curvature space. 
Second, let us construct gauge-invariant variables from coefficients of
the harmonics expansion. This procedure is done by analyzing the gauge
transformation of the coefficients and taking gauge-invariant linear
combinations of them. 
Thirdly, we can solve the constraint equations to obtain master
variables from the gauge-invariant variables. The constraint equations
are, of course, a part of $D$-dimensional perturbed Einstein equation. 
Finally, we can rewrite the remaining components of Einstein equation in
terms of the master variables to obtain master equations.

The master equation in general $D$-dimensions was first obtained in 
ref.\cite{Mukohyama2000b} and confirmed in ref.~\cite{KIS}\footnote{
In the latter paper, they extended the master equation of vector and 
tensor perturbations to more general background without maximal
symmetry.}. The master equations for generic values of the
$(D-2)$-dimensional momentum ${\bf k}$ (eg. ${\bf k}\ne 0$ for the $K=0$
case) are of the following form. 
%
\begin{equation}
 r^{\alpha+\beta}\nabla^a[r^{-\alpha}\nabla_a(r^{-\beta}\Phi)]
        - ({\bf k}^2+\gamma K)r^{-2}\Phi = 0,
\end{equation}
where $\Phi$ represents one of master variables $\Phi_{(S)}$,
$\Phi_{(V)}$ or $F_{(T)}$, $\nabla_a$ is the $2$-dimensional covariant
derivative compatible with the metric $\gamma_{ab}$, and $K$ is the
curvature constant of the $(D-2)$-dimensional constant-curvature
space. Two equivalent sets of constants ($\alpha$, $\beta$, $\gamma$)
are listed in Table~\ref{table:alpha-beta-gamma}. Master equations for
some exceptional values of ${\bf k}$ can be found in
ref.~\cite{Mukohyama2000b}.

%
\begin{table}[hbt]
\caption{Two sets of values of ($\alpha$, $\beta$, $\gamma$)}
        \label{table:alpha-beta-gamma}
\begin{center}
\begin{tabular}{|c||c|c|c||c|c|c|} \hline
 $\Phi$ & $\alpha$ & $\beta$ & $\gamma$ 
        & $\alpha$ & $\beta$ & $\gamma$ \\ \hline
 $\Phi_{(S)}$ & $D-4$ & $1$ & $0$ 
        & $-(D-6)$ & $D-4$ & $2(D-5)$ \\
 $\Phi_{(V)}$ & $D-2$ & $0$ & $-(D-3)$ 
        & $-(D-4)$ & $D-3$ & $D-3$ \\
 $F_{(T)}$ & $D$ & $-(D-3)$ & $-2(D-2)$ 
        & $-(D-2)$ & $2$ & $2$ \\ \hline
\end{tabular}
\end{center}
\end{table}

In the next sections we consider the $K=0$ case only. In this case
the exceptional value of ${\bf k}$ is ${\bf k}=0$. For ${\bf k}=0$, the
corresponding perturbations have the plane symmetry and, thus, the
generalized Birkoff's 
theorem guarantees that the perturbed bulk geometry is a S-AdS
spacetime. Hence, the perturbation with ${\bf k}=0$ is actually
perturbation of the mass parameter of the S-AdS spacetime around the
pure AdS and can be understood as dark radiation on the
brane~\cite{Mukohyama2000a}. Hence, we shall concentrate on
perturbations with non-zero ${\bf k}$. This treatment is, of course, 
justified by the fact that perturbations with different ${\bf k}$ are
decoupled from each other at the linearized level.

As an example, let us consider the case with $D=5$ and $K=0$. This
example is relevant for the $5$-dimensional brane world with spatially
flat background brane. 

In this case we decompose perturbations by harmonics on a
$3$-dimensional flat space. As shown in Table~\ref{table:variables}, we
have, for example, a gauge-invariant variable which transforms as a
$2$-dimensional scalar and a $3$-dimensional scalar. (In
Table~\ref{table:variables}, $Y$ is a scalar harmonics, $V_{(T)i}$ is a
transverse vector harmonics and $T_{(T)ij}$ is a transverse traceless
tensor harmonics.) We also have a variable which transforms as a
$2$-dimensional symmetric tensor and a $3$-dimensional scalar. Hence, we
have $(1+3)\times 1$ gauge-invariant degrees of freedom for
perturbations which transform as $3$-dimensional scalars. Similarly, we
have $2\times(3-1)$ and $1\times(6-3-1)$ gauge-invariant degrees of
freedom for perturbations which transform as $3$-dimensional transverse
vectors and transverse traceless tensors, respectively. Therefore, the
total number of gauge-invariant degrees of freedom is $10$. However, the
number of degrees of freedom of gravitons in $5$-dimensions is $5$. So,
we have too much gauge-invariant variables compared to the number of
gravitons.

On the other hand, after solving constraint equations, all we have are 
master variables which transforms as $2$-dimensional scalars. Hence, as
shown in Table~\ref{table:variables}, the total number of reduced
degrees of freedom is $5$. Therefore, the master variables concisely
describes gravitons in $5$-dimensions.

%
\begin{table}[hbt]
 \caption{Number of degrees of freedom}
 \label{table:variables}
 \begin{center}
  \begin{tabular}{|c||c|c|c|c|} \hline
 & $3$-D scalar & $3$-D T vector & $3$-D TT tensor  & $\#$ of variables
 \\ \hline
 $2$-D scalar & $FY\delta_{ij}$ & & $F_{(T)}T_{(T)ij}$ & \\
 $2$-D vector & & $F_aV_{(T)i}$ & & \\
 $2$-D tensor & $F_{ab}Y$ & & & \\ 
 \qquad (symmetric) & & & & \\
 & $(1+3)\times 1$ & $2\times (3-1)$ & 
 $1\times (6-3-1)$ & $10$ \\  \hline
 Master variables & $\Phi_{(S)}$ & $\Phi_{(V)}$ & $F_{(T)}$ & \\
 & $1\times 1$ & $1\times (3-1)$ & 
 $1\times (6-3-1)$ & $5$ \\ \hline
  \end{tabular}
 \end{center}
 \end{table}

\subsection{Perturbed junction condition}

Having the description of the bulk gravitational waves, what we have to
do is to investigate Israel junction condition~\cite{Israel}.

First, let us represent the world volume of a $(D-2)$-brane in a
$D$-dimensional spacetime by the parametric equations
%
\begin{equation}
 x^M = Z^M(y), 
\end{equation}
where $x^M$ ($M=0,\cdots,D-1$) and $y^{\mu}$ ($\mu=0,\cdots,D-2$) are 
$D$-dimensional coordinates and $D-1$ parameters, which play a role of
($D-1$)-dimensional coordinates on the brane world-volume. 
Next, let us consider perturbations of the functions $Z^M$ and the
$D$-dimensional metric $g_{MN}$. Then, we can calculate perturbations of
the induced metric and the extrinsic curvature of the hypersurface as
functions of $y^{\mu}$. Next, we can express the perturbed junction
condition in terms of these perturbations and matter perturbations on
the brane. Finally, by applying the perturbed junction condition to the
homogeneous isotropic background motion of the ($D-2$)-brane and
performing the harmonic expansion as in the previous subsections, we can
obtain junction conditions for gauge-invariant variables and master
variables.

The final expression can be found in refs.~\cite{Mukohyama2000c}. (See
also ref.~\cite{KIS}.) Here, I would only like to stress one important
aspect of the perturbed junction condition. 

First, since we are supposed to be living on the brane, physics in our 
world must not be affected by the following $D$-dimensional gauge
transformation in the higher dimensional bulk.
%
\begin{equation}
 x^M \to x^M + \xi^M(x).
\end{equation}
On the other hand, all observable quantities in our world must be
invariant under the following $(D-1)$-dimensional gauge transformation
on the brane. 
%
\begin{equation}
 y^{\mu} \to y^{\mu} + \zeta^{\mu}(y). 
\end{equation}
What is important here is that these two kinds of gauge transformations
are independent. One might expect that the ($D-1$)-gauge transformation
would be a part of the $D$-gauge transformation. In fact, as explicitly
shown in ref.~\cite{Mukohyama2000c}, it is not. Therefore, all physical 
quantities in our world on the brane must be invariant under these two 
independent gauge transformations. In particular, the junction condition
must, and actually can, be written in terms of doubly-gauge-invariant
variables only.


\section{Integro-differential equation}

We already have all basic equations. Namely, we have master equations in
the bulk and doubly-gauge-invariant junction condition. The next task we
have to do would be to simplify the basic equations to extract physics
from them. In this paper we consider scalar perturbations for $D=5$,
$K=0$.

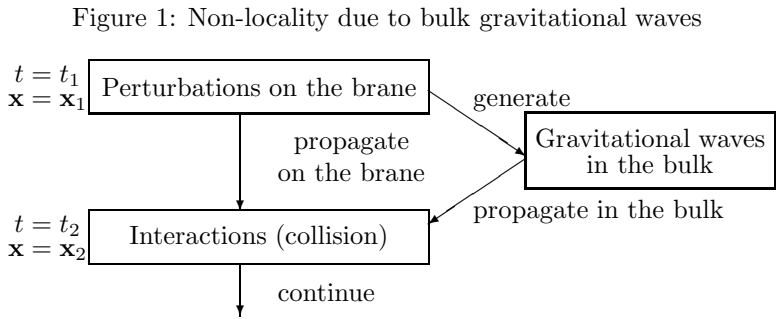
\begin{figure}[hbt]
 \caption{Non-locality due to bulk gravitational waves}
 \label{fig:non-locality}
 \setlength{\unitlength}{1mm}
 \begin{center}
  \begin{picture}(103,34)
   \put(12,27){\framebox(45,7){%
   Perturbations on the brane}}
   \put(70,17){\framebox(33,10){%
   \shortstack{Gravitational waves\\
   in the bulk}}}
   \put(12,7){\framebox(45,7){%
   Interactions (collision)}}
   \put(63,28){generate}
   \put(63,13){propagate in the bulk}
   \put(37,18){%
   \shortstack{propagate\\
   on the brane}}
   \put(37,2){continue}
   \put(0,28){
   \shortstack{%
   $t=t_1$\\
   ${\bf x}={\bf x}_1$}}
   \put(0,8){
   \shortstack{%
   $t=t_2$\\
   ${\bf x}={\bf x}_2$}}
   \put(57,30){\vector(3,-2){13}}
   \put(70,21){\vector(-3,-2){13}}
   \put(32,27){\vector(0,-1){13}}
   \put(32,7){\vector(0,-1){7}}
  \end{picture}
  \end{center}
\end{figure}

Before showing the result, let us think about what would be
expected. See figure~\ref{fig:non-locality}. First, let us consider a
perturbations localized on the brane. In other words, matter
perturbations. By definition, they propagate on the brane. However, at
the same time, they can generate gravitational waves. Gravitational
waves can, of course, propagate in the bulk spacetime, and may collide
with the brane at a spacetime point different from the spacetime point
at which the gravitational waves were produced ($t_1\ne t_2$, ${\bf
x}_1\ne {\bf x}_2$). When they collide with the brane, they should alter
evolution of perturbations localized on the brane. Hence, evolution of
perturbations localized on the brane should be non-local. It must, and 
actually can, be described by some integro-differential equations.

%

For scalar perturbations, the result of the simplification is three
differential equations and an integro-differential equation on the
brane~\cite{Mukohyama2001}. Two of the three differential equations can
be understood as the perturbed conservation equation, which is a general
consequence of the junction condition. These two are, of course, exactly
the same as the corresponding equations in the standard cosmology. The
other of the three differential equations differs from the corresponding
equation in the standard cosmology only by terms of order
$O(\rho/\lambda,p/\lambda)$, where $\rho$ and $p$ are energy density and
pressure of the background matter on the brane and $\lambda$ is the
tension of the brane. Therefore, that equation reduces to the equation
in the standard cosmology at low energy ($|\rho/\lambda|\ll 1$,
$|p/\lambda|\ll 1$).

The last of the four equations is an integro-differential equation of
the form 
%
\begin{equation}
 \tilde{R}(t) + \int dt'\ K(t,t')S(t') = 0, 
\end{equation}
if we assume that there is no gravitational waves coming from outside of
the Poincare patch of the bulk AdS~\footnote{For the modification by
gravitational waves coming from outside of the Poincare patch, or
initial gravitational waves, see ref.~\cite{Mukohyama2001}.}, where
$\tilde{R}(t)$ is a linear combination of gauge-invariant metric and
matter perturbations and their time-derivatives, $S(t)$ is a linear
combination of gauge-invariant matter perturbations, and the kernel
$K(t,t')$ is constructed from the retarded Green function of the master
equation for bulk gravitational waves. Here, we mention that all
physical quantities were constructed from coefficients of the
$3$-dimensional harmonic expansion and that the kernel $K(t,t')$ depends
on ${\bf k}^2$ quite non-trivially since the master equation depends on
${\bf k}^2$, where ${\bf k}$ is the $3$-dimensional momentum
vector. Thus, this equation indeed describes the non-locality due to
bulk gravitational waves: the matter perturbation $S(t')$ at the time
$t'$ generates gravitational waves and the gravitational waves, in turn,
affects the evolution of the perturbation $\tilde{R}(t)$ on the brane at
the different time $t$. The corresponding equation in the standard
cosmology is, of course, local and differs from $\tilde{R}=0$ only by a
term of order $O(\rho/\lambda,p/\lambda)$.

Therefore, if we consider a low energy regime, where 
$|\rho/\lambda|\ll 1$ and $|p/\lambda|\ll 1$, then the four equations
for scalar perturbations are almost the same as the corresponding
equations in the standard cosmology. The only difference is the
non-local term due to bulk gravitational waves. Since a large amount of
bulk gravitational waves can be produced at an early stage of the brane
universe and propagate in the bulk, there may be a possibility that the
non-local effect is significant even at a low-energy stage. Further
investigation is needed.


\end{document}